\documentclass[useAMS,usenatbib]{mn2e}
\usepackage{graphicx}
\usepackage{color}
\usepackage{amssymb}

\title{Hierarchical Bayesian analysis of the velocity power spectrum in supersonic turbulence}
\author[L. Konstandin, R. Shetty, P. Girichidis and R. S. Klessen]{
L. Konstandin$^{1}$\thanks{E-mail: email@lukaskonstandin.de}, R. Shetty$^{1}$, P. Girichidis$^{1,2}$ and R. S. Klessen$^{1,3,4}$\\
$^{1}$Universit\"at Heidelberg, Zentrum f\"ur Astronomie, Institut f\"ur Theoretische Astrophysik, Albert-\"Uberle-Str. 2, 69120 Heidelberg, Germany\\
$^{2}$Max-Planck-Institut f\"ur Astrophysik,  Karl-Schwarzschild-Str.~1, 85741 Garching, Germany\\
$^{3}$Kavli Institute for Particle Astrophysics and Cosmology, Stanford University,\\ \quad SLAC National Accelerator Laboratory, Menlo Park, CA 94025, USA\\
$^{4}$Department of Astronomy and Astrophysics, University of California, 1156 High Street, Santa Cruz, CA 95064, USA}
\begin{document}
\date{Accepted 01.April}
\pagerange{\pageref{firstpage}--\pageref{lastpage}} \pubyear{2014}
\maketitle
\label{firstpage}

%\input{/home/lkonstandin/Dokumente/Publications/Journal_abbrev_mnras}

%%%%%%%%%%%%%%%
%%% ABSTRACT  %
%%%%%%%%%%%%%%%
%% Not longer than 200 words
\begin{abstract}
Turbulence is a dominant feature operating in gaseous flows across
nearly all scales in astrophysical environments. Accordingly, accurately
estimating the statistical properties of such flows is necessary for
developing a comprehensive understanding of turbulence.  We develop
and employ a hierarchical Bayesian fitting method to estimate the
parameters describing the scaling relationships of the velocity power
spectra of supersonic turbulence.  We demonstrate the accuracy and
other advantages of this technique compared with ordinary linear
regression methods.  Using synthetic power spectra, we show that the
Bayesian method provides accurate parameter and error estimates.
Commonly used normal linear regression methods can provide estimates
that fail to recover the underlying slopes, up to 70\% of the
instances, even when considering the $2\sigma$ uncertainties.
Additionally, we apply the Bayesian methods to analyse the statistical
properties of compressible turbulence in three-dimensional numerical
simulations.  We model driven, isothermal, turbulence with root mean
square Mach numbers in the highly supersonic regime
$\mathcal{M}\approx15$.  We study the influence of purely solenoidal
(divergence-free) and purely compressive (curl-free) forcing on the
scaling exponent of the power spectrum.  In simulations with
solenoidal forcing and $1024^3$ resolution, our results indicate that
there is no extended inertial range with a constant scaling exponent.
The bottleneck effect results in a curved power spectrum at all
wave numbers and is more pronounced in the transversal modes compared
with the longitudinal modes.  Therefore, this effect is stronger in
stationary turbulent flows driven by solenoidal forcing compared to
the compressive one. The longitudinal spectrum driven with compressive
forcing is the only spectrum with constant scaling exponent
$\zeta=-1.94 \pm 0.01$, corresponding to slightly shallower slopes
than the Burger's prediction.

\end{abstract}

%%%%%%%%%%%%%%%%%%%%%%
%%% NOW THE KEYWORDS %
%%%%%%%%%%%%%%%%%%%%%%
%% one and six keywords
%% check list :http://oxfordjournals.org/our_journals/mnrasl/for_authors/mnraskeywords.pdf

%%%%%%%%%%%%%%%%%%%%%%%
%%% BODY OF THE PAPER %
%%%%%%%%%%%%%%%%%%%%%%%
\section{Introduction}
\label{Introduction}

Turbulence is a critical component of gaseous flows on nearly all
scales, as it is intimately related to many physical properties of the
medium, such as the morphology, mixing characteristics, and thermal
structure.  Turbulence is known to play a strong if not dominant role
in a variety of systems, from terrestrial incompressible flows
(e.g.~combustion engines, aerodynamics) to highly supersonic
compressible flows often occurring in astrophysical environments.
Consequently, accurately characterising the statistical properties of
turbulence is necessary for developing a comprehensive understanding
of fluid dynamics across a wide range of environments.

The statistical properties of turbulence, such
as the power spectrum, may serve as diagnostics for distinguishing
between different models. In the astrophysical context, for instance,
these are analytical and numerical models
describing accretion disks in
protoplanetary systems \citep[see e.g.][]{Meschiarj2012}, the
dynamics of the interstellar medium relevant for star formation
\citep[see e.g.][]{MacLow2004, McKee2007}, the formation
of star clusters and galaxies \citep[][]{Hopkins2012a} and galaxy
evolution \citep[][]{Iannuzzi2012}. Turbulence theory is also
important in the description of the diffuse interstellar medium
\citep[][]{Elmegreen2004} and for galactic or protogalactic
dynamos \citep{Brandenburg2005, Schober2012}. 
Despite its impact across a range of disciplines, a comprehensive theoretical
understanding of compressible turbulence remains elusive.

One key assumption of the \citet{Kolmogorov1941b} theory describing incompressible turbulence is that the energy transfer rate from large to small spatial scales $\epsilon$
should be constant. With the definition of a velocity fluctuation $\delta u_{\ell}$ at a length scale $\ell$ and its dynamical time-scale $\tau_{\ell} = \ell /\delta u_{\ell}$
one obtains
\begin{equation}
 \epsilon \simeq \frac{\delta u_{\ell}^2} {\tau_{\ell}}  \Leftrightarrow \delta u_{\ell} \simeq (\epsilon \ell)^{1/3}\,.
 \label{eq:pheno_scaling}
\end{equation}
This indicates that the velocity fluctuations can be described by a scaling law in the so called \textit{inertial range}, where the energy transfer rate $\epsilon$ is constant and the
flow is not influenced by viscous damping or the energy injection mechanism. From the power law behaviour of the velocity fluctuations a scaling law of the Kolmogorov velocity power spectrum
$P(k)$ can be derived,
\begin{equation}
P(k) \propto \epsilon^{2/3}k^{-5/3}\,.
\label{eq:pheno_PowerSpec}
\end{equation}
The kinetic energy is injected on the large scales and cascades to
small scales through non-linear coupling, until viscous effects become
important with respect to the advective terms.  At this ``dissipation
scale'' viscous effects cannot be neglected any more and the kinetic
energy is converted into heat (i.e.~internal energy).  This
description has to be extended for compressible turbulence.  In the
incompressible case, scale locality in $k$-space is crucial for the
Richardson-Kolmogorov picture of a cascade with constant energy flux
through the scales \citep{Frisch1995}. In $k$-space non-local, inter-scale
processes arising in compressible turbulence via shock fronts yield a more complicated situation
and equations (\ref{eq:pheno_scaling}) and (\ref{eq:pheno_PowerSpec}) have to be modified under these 
conditions.
The varying density field and the complex interplay between the density/pressure distribution with the velocity field have to be taken into account in a supersonic, compressible flow.
Note that it is not entirely clear which relation accurately describes the turbulent cascade.
Different approaches are used to expound the scaling of supersonic turbulence
\citep{Weizsaecker1951, Lighthill1955, Kida1990, Henriksen1991, Fleck1996, Kritsuk2007, Graham2010}.
For example, 
following the phenomenological arguments of \citet{Kolmogorov1941b}, \citet{Kritsuk2007} proposed a universal scaling behaviour for the power spectrum
of the quantity $\rho^{1/3}v$ \citep[see also][]{Galtier2011},
while others argue that the combination $\rho^{1/2} v$,
related to the kinetic energy, should be used instead \citep{Kida1990, Miura1995}.
Alternatively, one could look at the momentum transport with the unsymmetrical decomposition for the power spectrum in Fourier space $\widehat{\rho v}^* \widehat{v}$ \citep{Graham2010}.
%Following the phenomenological argument by \citet{Kolmogorov1941b} deriving equation (\ref{eq:pheno_scaling}) and (\ref{eq:pheno_PowerSpec}) \citet{Kritsuk2007} proposed a
%universal scaling behaviour for the power spectrum of the quantity $\rho^{1/3}v$, which was recently supported by an analytic derivation by \citet{Galtier2011}.}
%Therefore, an accurate measure of the scaling exponents is necessary for gaining a comprehensive understanding of supersonic turbulence \citep[e.g.~][]{Kaneda2003, Kritsuk2007, Lemaster2009, Federrath2010, Federrath2013}.

The theoretical predictions for the scaling exponents span only a
small range from $-5/3$ in the incompressible \citet{Kolmogorov1941b}
case, over $-2$ in the shock dominated \citet{Burgers1948} case and up
to $-19/9$ in the more recent theory of compressible turbulence of
\citet{Galtier2011} for the $\rho^{1/3}v$ spectrum.  Therefore, a high
precision measurement, as well as exact error estimates are needed to
distinguish between these model predictions.

Numerical simulations provide a viable avenue for measuring the
statistical properties of turbulent flows, and, by extension, testing
theoretical descriptions.  It is common practise to employ normal
$\chi^2$-based regression methods to estimate the scaling exponent of
the power spectrum of numerical simulations. Systematic errors, such
as the influence of the chosen fitting range, are normally not
explicitly treated. In this paper, we explore how common fitting
methods, and the associated assumptions, affect the resulting
parameter estimates.  We develop and compare a hierarchical Bayesian
technique with ordinary fitting methods, with the goal of quantifying
how well the power spectrum in numerical simulations follows an exact
power law. We focus here on the description of the methods and a
comparison with other methods\footnote{We use code written in the R programming language for our statistical analysis. It is available by sending a request to \textit{email@lukaskonstandin.de} .}.

Bayesian inference has the advantage that uncertainties in the data
are rigorously and self-consistently treated \citep[e.g.][]{Kelly2007,
  Gelman2004}. Additionally, Bayesian methods are well suited for
hierarchical problems, where different datasets, such as individual
snapshots, can be analysed simultaneously, providing parameter
estimates of both the individuals as well as for the whole
population. In astrophysics, Bayesian methods have been developed for
analysing observational data, such as turbulence in the ISM
\citep{Shetty2012}, analysis of dust extinction \citep{Foster2013} and
spectral energy distributions \citep{Kelly2012}.  Here, we apply a
general hierarchical model for the statistical analyses of turbulence
in numerical simulations. We demonstrate that the Bayesian method has
important advantages, including accurate parameter estimation, over
traditional non-hierarchical $\chi^2$-based methods.

The paper is organised as follows: Section
\ref{sec:Simulation_and_Methods} provides a description of the
simulations and the calculation of the spectra.  We also discuss the
caveats of ordinary fitting routines, explain our implementation of a
hierarchical Bayesian model and demonstrate its advantages on
synthetic data.  In Section \ref{sec:Results} we apply the Bayesian
model on the simulation data and interpret the results.
In the last Section we conclude and summarise our findings.\\

\section{Simulations and Methods}
\label{sec:Simulation_and_Methods}

\subsection{Properties of the simulations}
To model the dynamics of a turbulent gaseous flow, we solve the
equations of hydrodynamics, consisting of the continuity equation and the Euler equation with a stochastic forcing
term $\mathbf{F}$ per unit mass:
\begin{equation}
 \frac{\partial \rho}{\partial t} +(\textbf{v} \cdot \nabla)\rho=-\rho \nabla \cdot \textbf{v} \,,
\label{eq:continuity}
\end{equation}
\begin{equation}
\frac{\partial \textbf{v}}{\partial t} +(\textbf{v} \cdot \nabla)\textbf{v}=-\frac{ \nabla p}{\rho} + \textbf{F} \, ,
\label{eq:Euler}
\end{equation}
Here, $\rho$ denotes the mass density, $\textbf{v}$ the velocity
field, and $p$ the pressure.
Observations indicate that the dense
interstellar medium and molecular clouds behave as an isothermal flow
due to efficient cooling processes \citep{Elmegreen2004}. Accordingly,
we simulate with equation (\ref{eq:continuity}) and (\ref{eq:Euler}) an isothermal medium throughout this study
such that $p=\rho\mathrm{c_{\mathrm{s}}}^2$, with the sound speed $\mathrm{c_{\mathrm{s}}}$.

We employ the FLASH4 \citep{Fryxell2000, Dubey2008} code to solve the
set of partial differential equations (\ref{eq:continuity}) and
(\ref{eq:Euler}). We use the HLL5R solver \citep{Waagan2011}
on a uniform three-dimensional grid. To distinguish between physical and numerical effects, we run simulations with $512^3$, and $1024^3$ grid cells.

We compute the random forcing field $\textbf{F}$ in Fourier space as described by \citep{Schmidt2009, Federrath2010},
\begin{equation}
 \mathrm{d}\widehat{\textbf{F}}(\textbf{k}, t)=F_0(\textbf{k}, T_{\mathrm{ac}})\mathcal{P}^\zeta(\textbf{k})
 \frac{\mathrm{d}\textbf{W}(t)}{T_{\mathrm{ac}}}-\widehat{\textbf{F}}(\textbf{k}, t)\frac{\mathrm{d}t}{T_{\mathrm{ac}}}\,,
\label{eq:Forcing}
\end{equation}
where the $\mathrm{d}\textbf{W}(t)$ is a three-dimensional Gaussian random increment with zero mean and standard deviation $\mathrm{d}t$.
$\mathcal{P}^\zeta(\textbf{k})$ is a projection tensor in Fourier space as a function of the wave number $\textbf{k}$. In index notation, this operator is
\begin{equation}
{\mathcal{P}^\zeta}_{ij}(\textbf{k})=\zeta {\mathcal{P}^\bot}_{ij}(\textbf{k})+(1-\zeta) {\mathcal{P}^\|}_{ij}(\textbf{k})\,,
\label{eq:Decompos}
\end{equation}
where $\mathcal{P}^\bot=\delta_{ij}-k_ik_j/k^2$ and
$\mathcal{P}^\|=k_ik_j/k^2$ are fully solenoidal and compressive
projection operators, respectively, and $i,\,j$ are $\in[x,\,y,\,z]$.
The forcing has a finite autocorrelation time scale, $T_{ac}$, so that
it is smooth in space and time.
The forcing amplitude $F_0(\textbf{k})$ is a three-dimensional power-law function.
The forcing only occurs on the large (integral) scales $1 \leqslant
\left|\textbf{k}\right| \leqslant 2$, peaking at $k=1$, which
corresponds to the box size $L$, as we measure $k$ in units of
$2\pi/L$. The autocorrelation time-scale of the forcing algorithm is
set equal to the dynamical time-scale
$T_{ac}=T=L/(2\mathrm{c_{\mathrm{s}}}\overline{\mathcal{M}})$ and we
adjust the amplitude of the forcing field, such that the root mean
square Mach number is $\overline{\mathcal{M}} \approx15$.  As one of
our goals is to study the influence of the forcing scheme, we use the
projection tensor in Fourier space to get a purely solenoidal
(divergence-free, $\nabla \cdot \textbf{F} =0$) and a purely
compressive (curl-free, $\nabla \times \mathbf{F}=0$) vector field.

We start with homogeneously distributed gas at rest and let it evolve for $\approx 15\,T$ dynamical time scales.
The physical quantities in the simulations
are scale-free so that we define $L=1$, the mean mass-density
$ \left<\rho \right> = 1$ and $c_{\mathrm{s}} = 1$. We store the relevant quantities every $0.1\,T$ and
the fluid reaches the equilibrium state after about three turbulent crossing times, so that we have $121$ time snapshots in the state of fully developed turbulence.

\subsection{The Fourier spectra}
The Fourier spectrum of the velocity field is defined as
\begin{equation}
 \mathcal{P}(k)\mathrm{d}k =  4\pi k^2 \hat{\mathbf{v}}(k) \cdot \hat{\mathbf{v}}^*(k)\,\mathrm{d}k \, ,
\end{equation}
where $\hat{\mathbf{v}}$ is the Fourier-transformed velocity field and $\hat{\mathbf{v}}^*$ its complex conjugate.
With this definition the integral over the whole
$k$-range corresponds to the square of the Mach number,
\begin{equation}
\mathcal{M}^2 = \int_0^{\infty}\mathcal{P}(k) \mathrm{d}k\,,
\end{equation}
and the zero'th mode contains the averaged velocity field for velocity components $i\in {x,y,z}$,
\begin{equation}
\left< v_i\right> = \mathcal{P}_i(0)=\frac{1}{L^3}\int_0^{\infty} v_i(\textbf{r})\,\mathrm{d}^3\textbf{r}\,.
\end{equation}
In addition to the above
mentioned \textit{total} spectrum, we calculate the spectra of the
decomposed velocity field using the same decomposition as we use for
the forcing field (\ref{eq:Decompos}) and refer to these as the
\textit{longitudinal} and the \textit{transverse} spectra for the
curl-free and divergence-free velocity components, respectively.

We focus in our discussion on the analysis of the velocity power spectrum as an example.
It is the easiest and most commonly used statistical measure to describe turbulent flows.
We note that our conclusions about the fitting range and the comparison between our hierarchical Bayesian approach
and the standard linear regression methods hold for the distribution of other quantities as well.

\subsection{Caveats of the fitting method}
\label{subsec:caveats}
In practice, when analysing numerical simulations the scaling exponent
is often measured by linear regression in a log-log plot of the
time-averaged power spectrum, or on a $k^{5/3}$ or $k^2$ compensated
spectrum \citep[e.g.][]{Kaneda2003, Kritsuk2007, Lemaster2009,
  Federrath2010}.  We describe in the following four common
assumptions/methods that lead either to inaccurate scaling parameter
estimates, or to complications in interpreting the results.

\begin{figure}
\includegraphics[width=1 \linewidth]{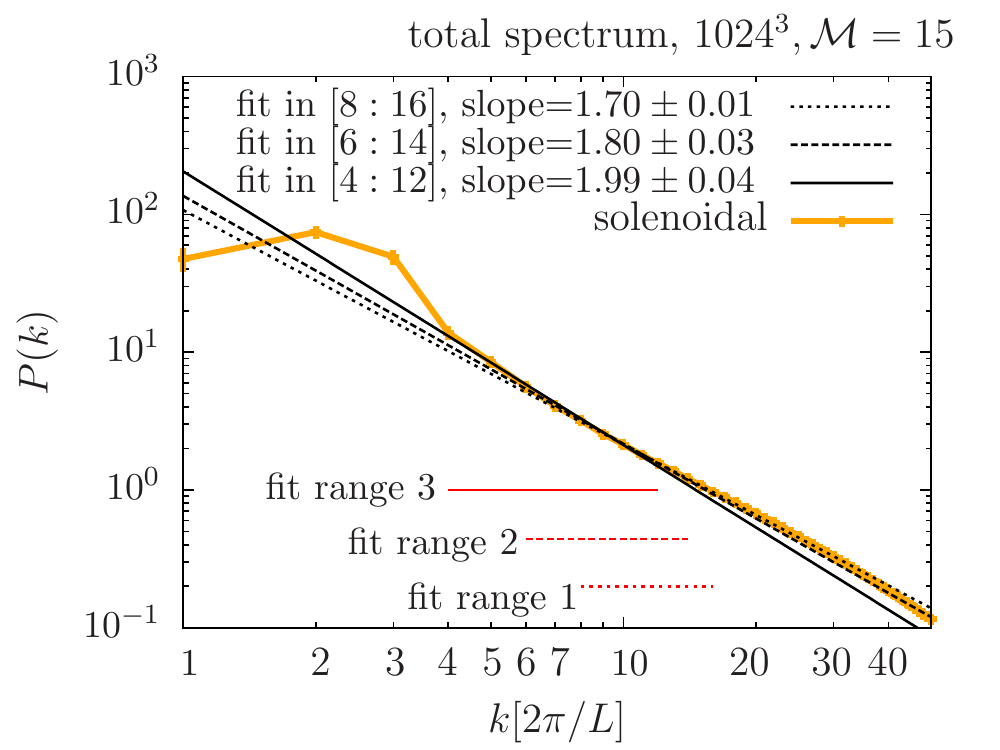}
\caption{Time-averaged power spectrum of the simulation with $1024^3$ grid cells and solenoidal forcing (orange points and error bars) and three different fits (black
 solid, dashed and dotted lines). The error bars correspond to the $1\sigma$ time variation of the power spectrum.
 All fits seem to describe the behaviour of the data in different $k$ ranges.}
\label{fig:Caveats}
\end{figure}

First, in a doubly logarithmic plot it is often difficult to verify if
the best-fit regression line accurately reproduces the data.  Many
functions may appear to follow a power law in a doubly logarithmic
plot.  For example, if the scaling exponent varies slightly with $k$,
a simple linear regression in log-space often does not reveal such
fluctuations. To demonstrate this caveat we perform three fits in
slightly varying ranges on the simulation with solenoidal forcing and
$1024^3$ grid cells (Figure \ref{fig:Caveats}).  The resulting fitting
parameters are listed in the figure.
All fits indicate a power law behaviour in the range $5 \lesssim k \lesssim 15$, although the measured slopes change significantly.
The reason for this is the curved
behaviour of the power spectrum, which does not follow a power law
over an extended range, as we will further discuss in Section
\ref{sec:Results}.  Hence, a qualitative validation whether the fit
can reproduce the measured data is needed.

Second, the $k$ extent of the inertial range is not known a priori.
The above example demonstrates the influence of the chosen fitting
range. It shows that the estimated slopes strongly depend on the
extent in $k$, because the power-spectrum slope of the data is not
necessarily constant in $k$.  Depending on the data, the measured
error of an ordinary linear regression method is very small and does
not describe the intrinsic uncertainty of the data (see error estimate
in figure \ref{fig:Caveats} and also Section \ref{sec:synthetic}), so
that this cannot be used to verify the quality of the chosen fitting
range.  In this case, an unbiased estimate of the inertial range is
very difficult to obtain.

Third, a key assumption in a $\chi^2$ linear regression is that the
uncertainties are independently and normally distributed.  The common
practice of fitting in log space implicitly assumes that the
uncertainties are normally distributed in log space. Usually the power
spectra are averaged to minimise their time-dependence and to reduce
the uncertainties and the scatter. However, averaging data also
assumes that the uncertainties of the data are Gaussian or at least
symmetrically distributed. Hence, both methods are based on the
assumption of a Gaussian/symmetric scatter, but for the linear space
as well as for the log space. Therefore, performing the averaging in
linear space and the $\chi^2$ fitting in log space is not consistent
and violates this underlying assumption.

Finally, information such as the time variation and the intrinsic
scatter contained in the data may be neglected when averaging data.
Hierarchical models exploit all the information in the data,
simultaneously estimating model parameters on multiple levels.  In the
next Section we introduce a hierarchical Bayesian method to account
for these issues for analysing the turbulent power spectra of
numerical simulations.

\subsection{Hierarchical Bayesian inference}
\label{sec:Bayesian_model}
To address the issues described above, we develop a hierarchical
Bayesian fitting method.  Hierarchical\footnote{Hierarchical modelling
  is often referred to as ``multi-level'' or ``random-effects''
  modelling \citep{Gelman2007}.}  modelling provides significant
advantages when the dataset is naturally structured into two or more
groups.  For instance, the hydrodynamic simulations provide spatial
information of all relevant quantities, such as the fluid densities
and velocities, at a series of snapshots in time.  The data is
therefore structured into temporal groups.  We can assess the
variation in the spectrum by analysing the datasets on the individual
time-level, as well as estimate the parameters of the mean
spectrum. Bayesian methods are well suited for estimating model
parameters on multiple levels in a hierarchical model.

With Bayes' theorem the probability $\mathcal{P}$ of a set of parameters $\bmath{\theta}$ given the
observed data $\textbf{D}$ can be calculated
\begin{equation}
 \mathcal{P}(\bmath{\theta} | \textbf{D}) \propto \mathcal{P}(\textbf{D}|\bmath{\theta})\mathcal{P}(\bmath{\theta})\,,
\end{equation}
where $\mathcal{P}(\textbf{D}|\bmath{\theta})$ is the probability of the set of data
 $\textbf{D}$ given the set of parameters $\bmath{\theta}$, known as the likelihood function $\mathcal{L}(\textbf{D}|\bmath{\theta})$.
$\mathcal{P}(\bmath{\theta})$ is referred to as the prior and is the probability of the set of parameters.
We will define $\bmath{\theta}$ in detail below.
The outcome of Bayesian inference is the probability of the model parameters $\bmath{\theta}$ given the data $\textbf{D}$
and is called posterior distribution.
To evaluate the posterior, we perform a Markov Chain Monte Carlo (MCMC) sampling of $\bmath{\theta}$ for constructing the product of the prior and likelihood.
The result of the Bayesian inference, the posterior, is the joint
probability distribution of the parameters.
The errors in each measured quantity are assumed to be drawn from some
a priori defined distributions described by one of the parameters.
For a detailed description of the Bayesian inference method, we refer
the reader to the standard textbooks about statistical methods
\citep{Gelman2004, Kruschke2011, Wakefield2013}.

In the following we will describe the construction of the Bayesian model,
using the standard statistical notation.
We describe how quantities are conditionally related, such that $x|y$ refers to a variable $x$ given a value of $y$.
Characterising values and their distribution, like $x|\mu,\, \sigma^2 \sim \mathcal{N}\left(\mu,\, \sigma^2 \right)$ denotes
that $x$ is drawn from a normal distribution
\begin{equation}
\mathcal{N}\left(x|\mu,\,\sigma^2 \right) = \frac{1}{\sqrt{2\pi}\sigma}\exp{\left(\frac{-(x-\mu)^2}{2\sigma^2} \right)}\,,
\end{equation}
given the mean value $\mu$ and the variance $\sigma^2$.
We also employ gamma distributions
\begin{equation}
\mathcal{G}(x|s,\,r) = \frac{r^s}{\Gamma(s)} x^{s-1}\exp{\left( -r x\right)}\,,
\end{equation}
for the inverse of the variance with $s$ and $r$ being the shape and rate parameters, respectively, and $\Gamma$ the gamma function.
Before performing the fit we standardise the data, i.e.~we transform it with
\begin{equation}
 \widetilde{y} \equiv \frac{y-\mu_y}{\sigma_y} \,,\quad \widetilde{x} \equiv \frac{x-\mu_x}{\sigma_x}\,,
\label{eq:standardize}
\end{equation}
where $\mu$ and $\sigma$ indicate the mean and the standard deviation. This has the advantages that
we know exactly the parameter range over which we have to sample with the ``hyperpriors'' (see definition further below).

In a Bayesian model all quantities are drawn from some a priori defined
distributions. Therefore, we assume that the velocity power spectrum
$P(k, t)$ follows a power law, i.e.~a linear function in log-log space. Additionally, we include
a scatter term $\delta_s(k_i,t_j)$, which measures the deviations from a perfect power law,
\begin{equation}
\log{P(k_i, t_j)} = A(t_j) + \log{k_i}*{\zeta(t_j)} +\delta_s(k_i,t_j)\, .
\label{eq:Power_law}
\end{equation}
This equation describes the relationship
between the parameters on the individual level in the hierarchy.
The intercept $A(t_j)$, the power-law index $\zeta(t_j)$ and
the scatter $\delta_s(k_i,\,t_j)$ of each
individual time snapshot $t_j$ must be drawn from the prior conditional probability distributions
\begin{equation}
 A(t_j) | \overline{A}, \overline{\sigma^2_A} \sim \mathcal{N}\left( \overline{A}, \overline{\sigma^2_A} \right)\,,
\end{equation}
\begin{equation}
\zeta(t_j) | \overline{\zeta}, \overline{\sigma^2_{\zeta}} \sim \mathcal{N}\left( \overline{\zeta}, \overline{\sigma^2_{\zeta}} \right)\,,
\label{eq:bay_slope}
\end{equation}
\begin{equation}
\delta_s(k_i,\,t_j) | \sigma^2_{\Delta}(t_j) \sim \mathcal{N}\left( 0, \sigma^2_{\Delta}(t_j) \right)\,,
\end{equation}
\begin{equation}
1/ \sigma^2_{\Delta}(t_j) | \overline{s}, \overline{r}  \sim \mathcal{G} \left( \overline{s}, \overline{r} \right)\,.
\label{eq:Bayes_model}
\end{equation}
The model uses normal distributions for the slope, intercept and the scatter and a gamma distribution for the inverse of the variance
of the scatter term. The inverse of the variance is also called precision.
We chose a gamma distribution for the precision of the scatter to have a really broad prior,
as we would like to rely on the data and not the priors.

Those quantities that depend on $t_j$ refer to individual time frames. For instance, $\zeta(t_j)$ is the slope of the time snapshot $t_j$
whereas $\overline{\zeta}$ refers to the group slope of the whole dataset.
The fitting results of each relationship above depend on quantities from the higher group level of the hierarchy,
i.e.~describe the time-averaged behaviour of the power spectrum. The prior assumptions for this final level, which are called
``hyperpriors'', are
\begin{equation}
\overline{A} \sim \mathcal{N}\left( 0, 10 \right)\,,
\label{eq:prior_A}
\end{equation}
\begin{equation}
\overline{\zeta} \sim \mathcal{N}\left( 0, 100 \right)\,,
\label{eq:prior_zeta}
\end{equation}
\begin{equation}
1/\overline{\sigma^2_A} \sim \mathcal{G} \left( 0.1, 0.1 \right)\,,
\label{eq:prior_p}
\end{equation}
\begin{equation}
1/\overline{\sigma^2_{\zeta}} \sim \mathcal{G}  \left( 0.1, 0.1 \right)\,,
\end{equation}
\begin{equation}
\overline{s} | m, d = m^2 / d^2\,,
\label{eq:s}
\end{equation}
\begin{equation}
\overline{r} | m, d = m / d^2\,,
\end{equation}
\begin{equation}
m \sim \mathcal{G}  \left(1, 0.1 \right)\,,
\end{equation}
\begin{equation}
d \sim \mathcal{G}  \left(1, 0.1 \right)\,.
\label{eq:prior_d}
\end{equation}
The model contains the mean, $m$, and standard deviation, $d$, of the scatter term,
as they are more intuitive than the shape and rate parameters of the gamma distribution.
The mean $\mu$ and variance $\sigma^2$ of a gamma distribution with shape $s$ and rate $r$ is defined as,
$\mu=s/r$ and $\sigma=s/r^2$, respectively.

Recall that we normalize the data with equation
(\ref{eq:standardize}), so that the averaged intercept is zero and the
standardized slope is just the correlation $corr(x,y) \in [-1:1]$.
Therefore, we have broad ``hyperpriors'' in the model, such that the
fixed values in (\ref{eq:prior_A})-(\ref{eq:prior_d}), e.g.~the group
slope, is drawn from a normal distribution with mean $\mu=0$ and
$\sigma^2=100$. All values in (\ref{eq:prior_A})-(\ref{eq:prior_d}) do
affect the number of samples until the Markov Chain Monte Carlo method
converges, but assuming sufficient sampling and that the ``true''
values lie inside the priors, they do not affect the end results of
the Bayesian inference.  For a more detailed description of the
construction of a Bayesian model we refer the reader to standard
textbooks of statistical data analysis \citep{Gelman2004,
  Kruschke2011, Wakefield2013} or recent publications using similar
models \citep{Kelly2007, Shetty2013, Shetty2014}.

In summary, this Bayesian method explicitly treats the common fitting
issues mentioned in the last Section. That is, variations of the
scaling exponents with time yield a larger variance of the group
slope. Fluctuations of the scaling exponents with $k$ increase the
group scatter $\sigma^2_{\Delta}(t_j)$.  Variations and uncertainties
of the measured data are also treated self-consistently.  Both
individual and also the global parameters are estimated
simultaneously, avoiding any data-averaging. Since defining a fitting
range introduces a large uncertainty, we test the Bayesian model on
synthetic data in the next Section, where we fit over a $k$ range of
seven points to obtain the ``local`` slope of the power spectrum.

\subsection{Test with synthetic data}
\label{sec:synthetic}

\begin{figure*}
\includegraphics[width=1 \linewidth]{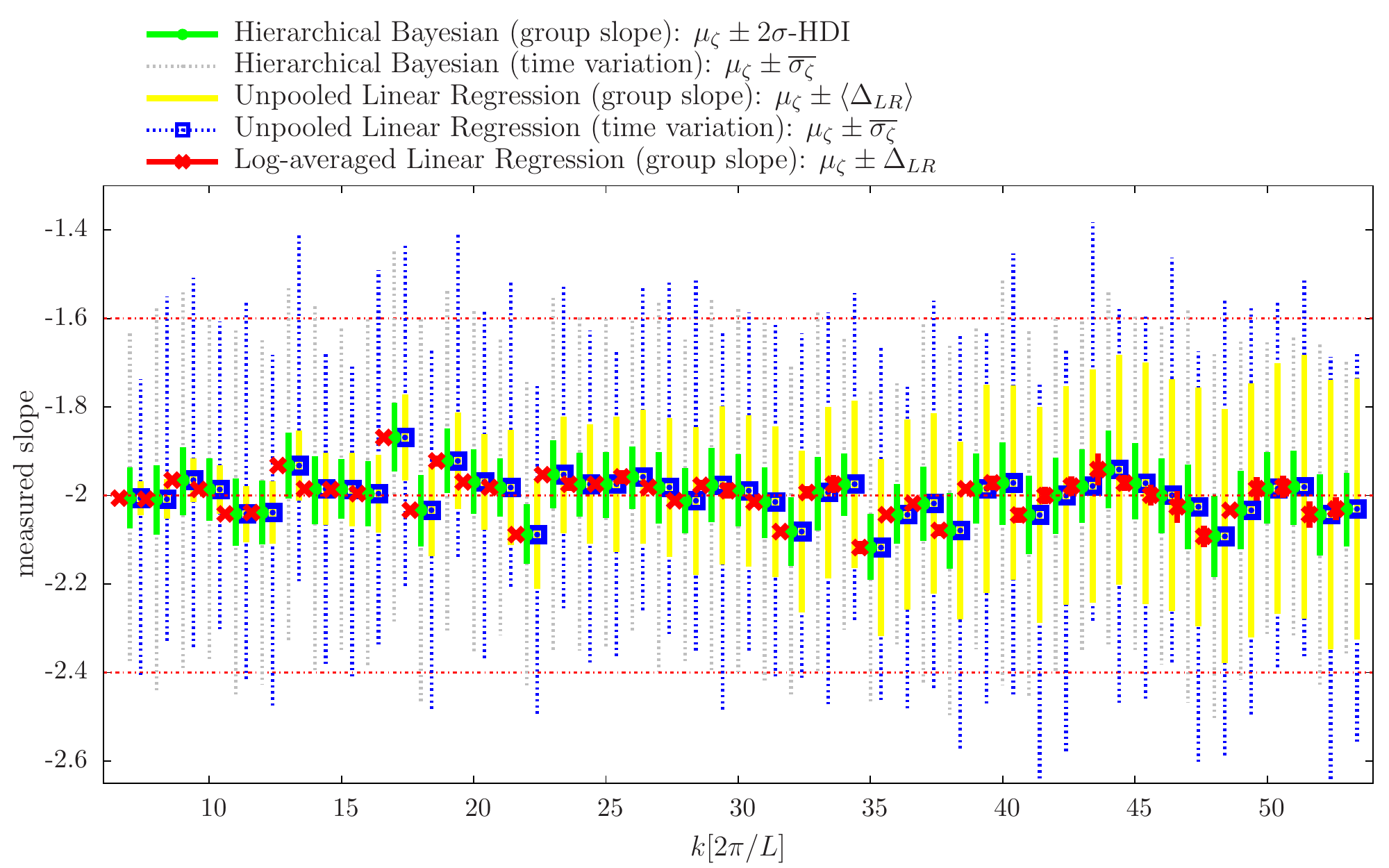}
\caption{Test on synthetic data with a slope of $-2$ and time
  variation of $0.4$ (indicated by the horizontal dashed-dotted
  lines).  The ordinate indicates estimates of the group slope, from
  three different methods, over an extent of $\Delta k = 6$, plotted
  with the centre value of $k$ on the abscissa .  We compare the
  Hierarchical Bayesian, an unpooled linear regression to mimic
  hierarchical modelling with ordinary linear regression, and an
  ordinary linear regression applied to the spectra averaged in
  log-space.  With the former two methods we estimate the variation of
  the slope with time (dashed thin lines) as well as the uncertainty
  of the group slope (thick solid lines).  The results of the
  log-averaged linear regression is shifted slightly to the left,
  whereas these of the unpooled method are slightly shifted to the
  right for clarity.  The creation of the synthetic data and the
  employed methods are discussed in more detail in Section
  \ref{sec:synthetic}.}
\label{fig:Synthetic_k}
\end{figure*}

We verify the hierarchical Bayesian model with synthetic data and compare it with
normal linear regression (LR) methods. We create a synthetic dataset with $121$
realisations according to the Bayesian model (equations
(\ref{eq:Power_law})$-$(\ref{eq:Bayes_model})), where the
group intercepts, the slopes and the scatter-precision follow
distributions with mean values of $(5, -2, 1000)$ and standard deviations of $(1, 0.4, 200)$.
These parameters for creating the synthetic data reflect the averaged behaviour of the measured power spectra in log-log space,
where we slightly overestimate the variation in time. The synthetic data are  distributed logarithmically on the x-axis instead of homogeneously distributed,
as we will apply the methods in log-log space.
Figure \ref{fig:Synthetic_k} shows the slope measured in a fitting range $\Delta k = 6$ (with seven points)
as a function of the point in the centre of the fitting range for different methods.
As we fixed the size of the fitting range in linear space
its width is decreasing with $k$ in log space, which we will discuss further below.
The hierarchical model rigorously accounts for a number of uncertainties. The posterior probability distribution function (PDF) contains the resulting fit parameters,
for both the group and the individuals.
For example, the width of the PDF, or highest density interval (HDI), of the group slope and intercept yields the range
in plausible parameters, considering the measurement uncertainty or insufficient statistics, caused by fitting only seven points.

Figure \ref{fig:Synthetic_k} shows estimates for two different parameters of the synthetic data.
The group slope of the spectra with a $2\sigma$-HDI uncertainty estimate, as well as the $1\sigma$ variation of the slopes with time without an uncertainty estimate.
The green circles correspond to the Bayesian measurement of the group slope $\overline{\zeta}$ (the mean value in equation \ref{eq:bay_slope} and \ref{eq:prior_zeta})
and its $2\sigma$-HDI interval (green, solid, thick lines), whereas the grey, dashed, thin lines quantify the variations of the slope in time using the maximum likelihood value of
the standard deviation $\overline{\sigma_{\zeta}}$ in equation (\ref{eq:bay_slope}) and (\ref{eq:prior_p}).
To mimic hierarchical modelling using a normal linear regression we perform a fit on each individual time realisation and collect
the slopes and error estimates in two histograms. The mean value of the resulting histogram with $121$ error estimates gives a measurement
of the averaged error of the fits (yellow, solid, thick lines). The mean value of the resulting histogram with $121$ slopes provides
an estimate of the group slope (blue squares) and its $1\sigma$-HDI measures the variation in time (blue, dashed, thin lines). 
We refer to this method as 'LR-unpooled' further below,
as it does not average the power spectra of the different time snapshots.
The red crosses correspond to a normal linear regression method
applied to the spectra averaged in log space.

All methods in figure \ref{fig:Synthetic_k} have a comparable accuracy
for estimating the maximum likelihood slope, which does not depend on
the scale in the shown range, whereas the error estimates are
significantly different.  With normal linear regression applied to the
log-averaged spectra, in nearly all cases ($71\%$) the true slope lies
outside the error interval (the red crosses are in most cases larger
than the uncertainty intervals).  Alternatively, the error estimates
of the unpooled linear regression (yellow, solid, thick) contain the
correct value in all but one case, and the Bayesian method (green,
solid, thick) contain the correct value in $92\%$ of all cases.  The
uncertainty interval of the unpooled linear regression should contain
the correct value in $68\% = 1\sigma$, as we calculate it with the
mean value of the histogram of the individual errors. It increases
systematically with $k$, which is due to an interplay of the
decreasing width of the fitting range with $k$ in log space and the
increasing importance of the scatter with $k$, making this method
impractical for a high precision measurement of the scaling exponent
of the power-spectrum. On the other hand, both the unpooled linear
regression model (blue, dashed, thin) as well as the hierarchical
Bayesian model (grey, dashed, thin) recover the variation with time of
the group slope of $\pm 0.4$.

Figure \ref{fig:Synthetic_k} indicates that the
regression method can have a major influence on the results, especially the error estimate, and should be chosen carefully.
The ordinary linear regression applied to the averaged spectrum stands out negatively, as its error estimate of the mean slope totally fails.
The implementation of a method to mimic hierarchical modelling using a normal $\chi^2$-linear regression
can recover the time variation of the group slopes, but its measurement of the averaged error between the individuals cannot be used to quantify the uncertainty of the
group slope, as it strongly depends on the scale $k$ and gets too large to distinguish between the different theoretical models.
This is caused by an interplay of two effects. First, as we assume a fixed distribution for the scatter the relative importance of the scatter increases with $k$, which
the unpooled linear regression cannot handle. Second, as we fix the fitting range in linear space, but fit in log-log space the effective width of the fitting range decreases with $k$,
influencing the error estimate for the unpooled linear regression method.
The Bayesian method, on the other hand, recovers all information about the slope with a high precision and valid error estimates.

\section{Results}
\label{sec:Results}
\begin{figure}
\includegraphics[width=1 \linewidth]{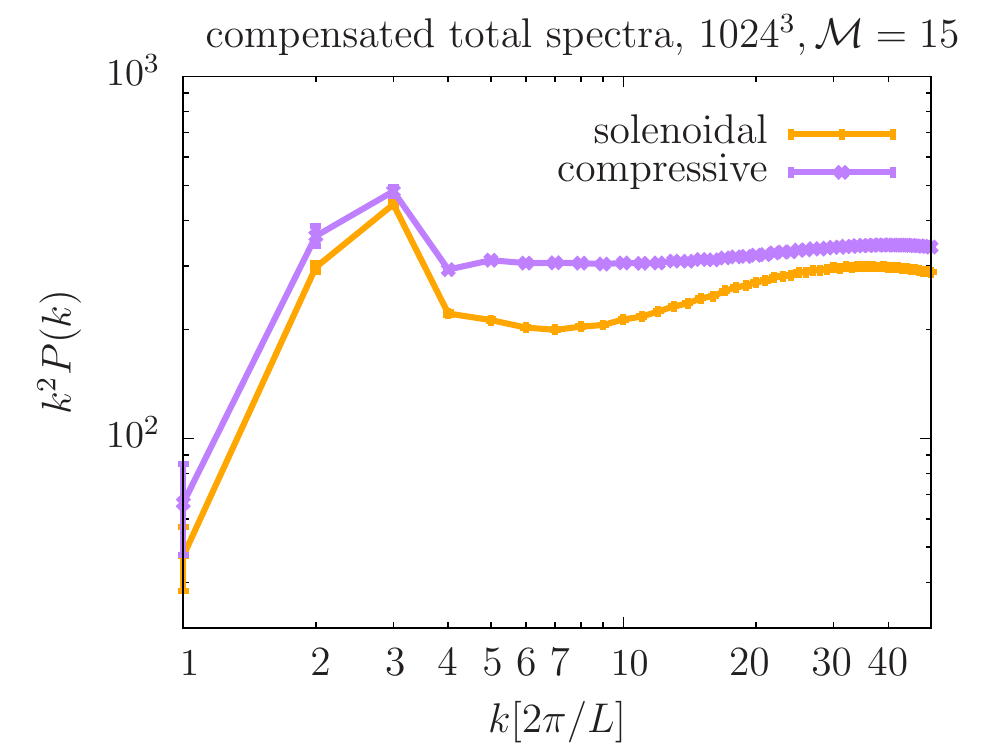}
\caption{Total spectra for solenoidal (orange) and compressive (purple) forcing, compensated with $k^2$, and $1024^3$ resolution.}
\label{fig:Spec_compensated}
\end{figure}

Figure \ref{fig:Spec_compensated} shows the total spectra for solenoidal (orange) and compressive (purple) forcing, compensated with $k^2$, and
for the simulation with $1024^3$ resolution. It clearly indicates that the compressive forcing yields a spectrum following the Burgers prediction over an extended range,
whereas the solenoidal forcing yields a curved spectrum. The bump of energy at intermediate scales $k \approx 20-40$ is caused by a phenomenon
normally known as the bottleneck effect \citep[e.g.~][]{Dobler2003, Schmidt2006, Donzis2010}. We will discuss its influence on the
spectra in detail further below using the Bayesian estimate of the scaling exponent.

\begin{figure}
\includegraphics[width=1 \linewidth]{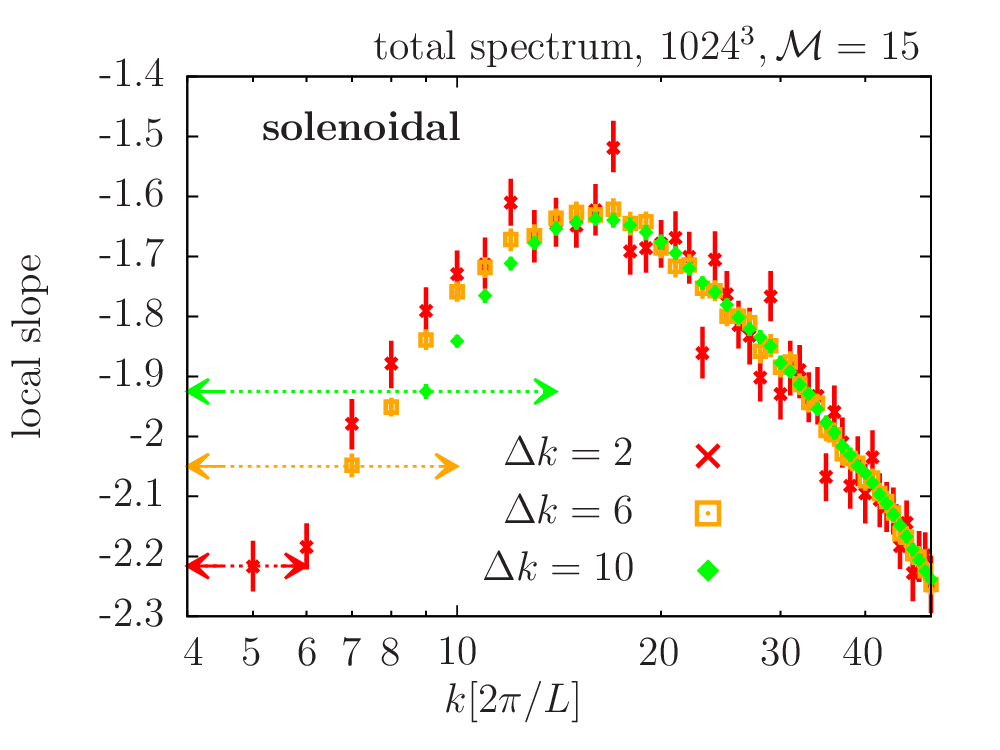}
\caption{Measured local group slope of the Bayesian method as a function of the window centre $k$ for three different fitting window sizes
$\Delta k = 2,\, 6, \, 10$ (red, orange, green) performed on the total spectrum of the simulation with $1024^3$ grid points and
solenoidal forcing.}
\label{fig:Fitting_range}
\end{figure}

Next, we test how the extent of the fitting range influences the
measured scaling exponents.  We do this on the measured spectra
instead of synthetic data and therefore use the simulation with
solenoidal forcing and $1024^3$ resolution.  Figure
\ref{fig:Fitting_range} shows the measured group slope
$\overline{\zeta}(k)$ as a function of the centre of the fitting range
$k$ for three different widths of the fitting range $\Delta
k=2,\,6,\,10$ (thereby including $3,\,7,\,11$ points).  Increasing the
fitting range decreases the uncertainty in the measured scaling
exponent.  It also averages the high-frequency scatter out, without
changing the global functionality on $k$.  On low $k$ values the
measurements with small fitting windows estimate steeper slopes. But
this is not a systematic error included by the small fitting ranges.
It can be explained by the changing slopes of the power spectra in the
given ranges. We indicate the $k$ ranges of the different fitting
windows on the first point of each measurement as a horizontal dashed
line. The fitting ranges of all measurements start at $k=4$, where the
forcing routine has no direct influence any more.  Figure
\ref{fig:Fitting_range} shows that the spectrum is strongly curved
with a steep area at low wave numbers and gets systematically shallower
with increasing $k$.  So the steep part at small scales $k\approx 5$
influences the first measurement of the $\Delta k=10$ curve at $k=9$
(first green measurement), whereas the measurement with $\Delta k =
2$ at $k=9$ is only influenced by the slope in $k \in (8:10)$ and is
therefore systematically shallower (fifth red point).  Figure
\ref{fig:Fitting_range} indicates that the scaling exponents of the
solenoidal run span the whole range of theoretical predictions in the
scale range $k \in (5 : 15)$.

\begin{figure*}
\includegraphics[width=1 \linewidth]{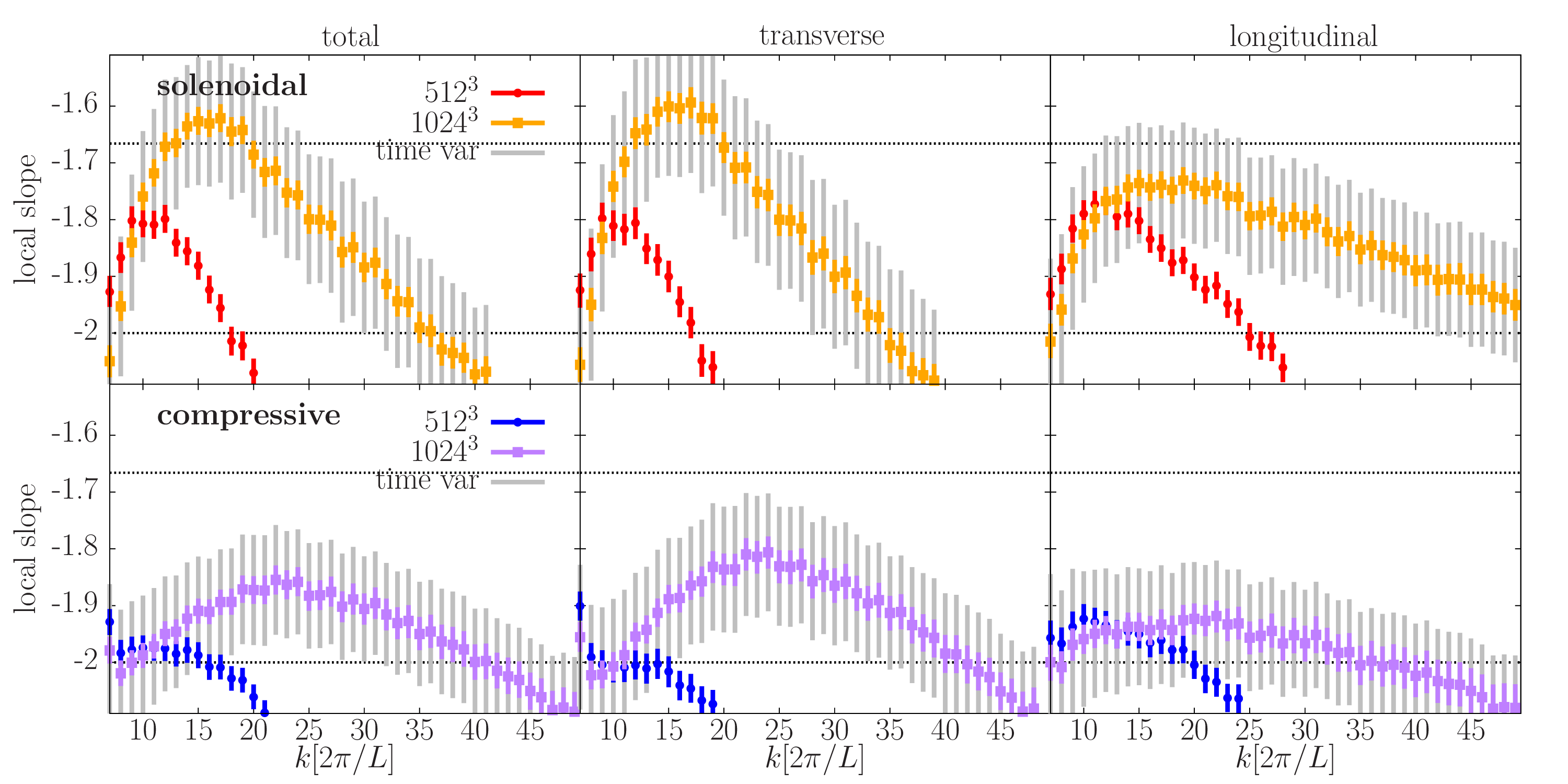}
\caption{Local group slope as a function of the centre of the fitting window $k$ with a size of $\Delta k =6$ applied to the total (left),
transverse (middle), and longitudinal (right) spectra of the simulations with $512^3,\,1024^3$ resolution (red/orange and blue/purple),
solenoidal (upper panels) and compressive (bottom panels) forcing. The grey error bars indicate the time variation of the slope at each $k$ for the $1024^3$ simulations.
The horizontal dotted lines indicate Kolmogorov $-5/3$ scaling and a Burgers $-2$ scaling behaviour.}
\label{fig:Localslope_modes}
\end{figure*}

Figure \ref{fig:Localslope_modes} shows the local group slope measured
with window size $\Delta k = 6$ as a function of the centre of the
fitting range $k$ for solenoidal (upper panels) and compressive
forcing mechanism (bottom panels), each for different resolutions
$512^3$ and $1024^3$ (red/orange and blue/purple, respectively), and
from left to right the local slope of the total, transverse and
longitudinal decomposed spectra.  The grey error bars indicate the
time variation of the slope at each $k$ only for the $1024^3$
simulations. As we measure $k$ in units of $2\pi /L$ with constant $L$
for different resolution, the spectra should overlap on the large
scales (low $k$). The spectra with $512^3$ and $1024^3$ resolution
deviate from each other already on the large scales, indicating that
they are not converged with resolution.  All spectra are curved in the
displayed range with a slope of $\approx -2$ at large scales close to
the forcing routine, a shallow area at intermediate scales, and
systematically decreasing slopes in the range, where the numerical
dissipation can no longer be neglected.  This "bump" is more
pronounced for the transverse spectra than for the longitudinal and is
still increasing with resolution.  Its peak appears for solenoidal
forcing on larger scales and with shallower slopes than for
compressive forcing. The longitudinal spectrum in the simulation with
compressive forcing is the only case with a constant slope over an
extended range $k \in (10:32)$, which corresponds to $102,\,32$ grid
cells.  Applying the Bayesian model to this range produces a group
slope $\overline{\zeta}= -1.94$ with the small $2\sigma$-HDI
$[-1.95:-1.93]$ and a standard deviation for the time variations
$\overline{\sigma_{\zeta}} = 0.04$.

\begin{figure}
\includegraphics[width=1 \linewidth]{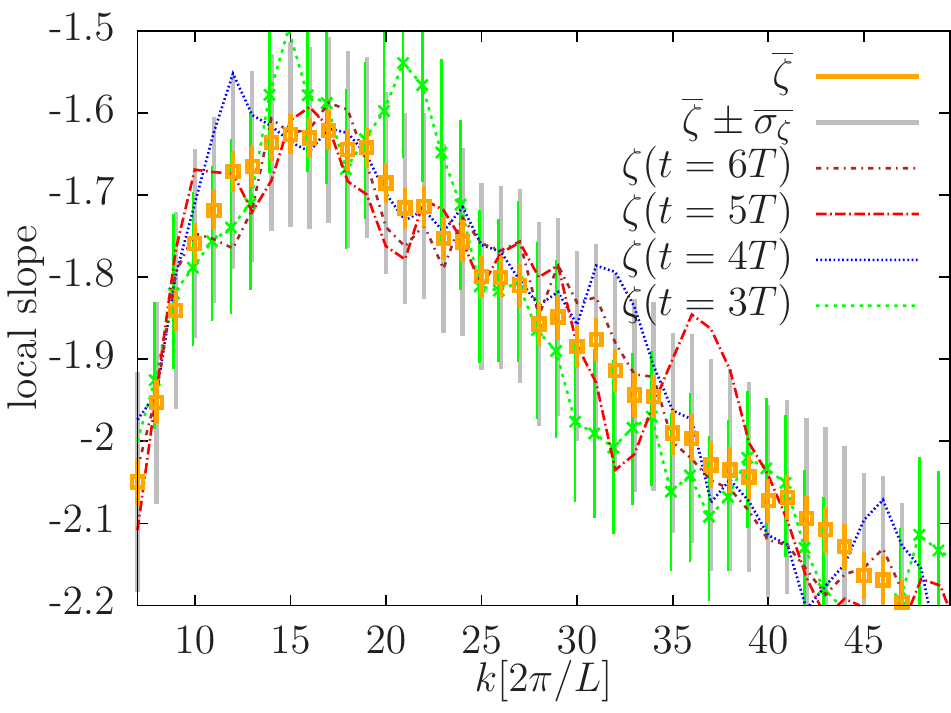}
\caption{Same as in the left, upper panel of figure
  \ref{fig:Localslope_modes}. In addition, we show the estimates of
  the slope in the individual times $t \in (3,\,4,\,5,\,6)[T]$ to
  illustrate the high time variation. We provide only the uncertainty
  interval of the $t=3[T]$ individual slope estimation and skip these
  for the other times for clarity.}
\label{fig:Local_slope_individuals}
\end{figure}
The simulation data indicate large temporal variations of the slopes $\zeta$ with variance $\overline{\sigma_{\zeta}} \approx 0.1-0.2$
(grey error bars in figure \ref{fig:Localslope_modes}) for a window size of $\Delta k = 6$, which is independent of the forcing,
$k$ scale, and the mode of the analysed spectra. Increasing the fitting range decreases the temporal fluctuations
(compare with the fit results $k \in (10:32)$ for the longitudinal spectrum and compressive forcing stated above).
Figure \ref{fig:Local_slope_individuals} shows the same as the left, upper panel of figure \ref{fig:Localslope_modes}, but in addition it provides the estimates of the individual slopes
at the times $t \in (3,\,4,\,5,\,6)[T]$ to illustrate the fluctuations of the slope $\zeta$ for different times.

\subsection{Discussion and interpretation}
The measurements show that the power spectra are curved and not converged at a resolution of $512^3$ and $1024^3$ grid cells.
An accumulation of kinetic energy just before the dissipation wave number is a phenomenon called ''bottleneck effect`` \citep[e.g.~][]{Dobler2003, Schmidt2006, Donzis2010}.
We interpret the bump in the slopes as an influence of a numerical, non-physical bottleneck effect for two reasons.
First, the height of the bump is still varying
with resolution. And second, we rely on numerical dissipation, which is in agreement with the studies of \citet{Lamorgese2005} showing that the bottleneck in the energy spectrum becomes more pronounced as the hyperviscosity index is increased.

The bottleneck effect peaks at $k=16,\,23$ for the $1024^3$ simulations with solenoidal and compressive forcing, respectively.
It is more pronounced in the transverse than in the longitudinal spectrum indicating that the dissipation of the transverse modes of the velocity field is fainter than that of the
longitudinal modes. Increasing the number of shocks in a simulation by changing the forcing modes from solenoidal to compressive at constant Mach number decreases the amplitude
of the bottleneck effect. We interpret this with the non-local energy flux through the scales introduced by shocks, which allows the flow to jump over a range of scales instead of
transporting it steadily through the scales. However, a detailed study of the energy fluxes of the different velocity modes is necessary to validate this interpretation.

The reason for the large fluctuations in the slope $\zeta$ measured at different times can be explained as follows.
Employing a constant forcing amplitude in (\ref{eq:Forcing}) fixes the resulting Mach number only in a statistical sense. The actual energy
and and momentum injection varies with time depending on the correlation of the density field and the forcing field.
If the forcing pattern overlaps by chance with a high density region, more energy gets injected, causing time fluctuations in the velocity field.
These are visible on the power spectra yielding the variations of the slopes with time.

\section{Conclusions}
We introduced a hierarchical Bayesian method for estimating the
scaling exponent of the velocity power spectrum.  We validated it using 
synthetic data and compared it with ordinary linear regression models
applied to the log averaged power spectrum and an unpooled linear
regression method to mimic hierarchical modelling.  We demonstrate
that the ordinary linear regression model, applied to the averaged
spectra, produces parameter estimates that fail to recover the
underlying slope in $\approx70\%$ of the measured points, within the $2\sigma$
uncertainties.  With the unpooled linear regression method the time
variation of the slope can be accurately recovered, but the error
estimate of the mean slope systemically increases with the scale $k$
up to $\approx 0.2$ at $k=30$, which spans basically the entire range of slopes predicted by theoretical models and can thus not be used to distinguish between them.
The hierarchical Bayesian method avoids
the caveats of the linear regression methods and can recover the
underlying mean behaviour of the power spectrum, its time variation,
as well as all errors and uncertainty estimates on these
quantities. Therefore, the Bayesian method provides more information,
and because of the correct error estimate, more robust parameter
estimates of the power spectrum.
Additionally, we implemented a routine to apply the hierarchical Bayesian method to fitting windows,
where we change the sizes and placements systematically, to estimate the uncertainty caused by defining a fitting range.
All improvements of the presented Bayesian method can also be achieved with the unpooled linear regression method besides the error estimate of the mean slope
and an estimate of the scatter.\\

To demonstrate the improvements of such an analysis we applied it to a "standard" simulation setup for analysing supersonic turbulence.
The simulations have $1024^3$ resolution, a large scale forcing field (decomposed in solenoidal and compressive modes), which accelerates the isothermal gas to
a root mean square Mach number of $\mathcal{M}\approx 15$. The grid based simulation code includes artificial 
numerical dissipation. Our findings are:\\
\begin{enumerate}
\item The resolution study with $512^3$ and $1024^3$ showed that the scaling exponents of the spectra are varying significantly with time and scale and are not converged with resolution.\\
\item Independent of the forcing mechanism, we can rule out with $2\sigma = 95\%$ certainty that neither the total, nor the transverse
spectra show an extended range where the power spectra have constant scaling exponents.
They start at $k=4$ with a slope of $\approx -2$  for solenoidal (compressive) forcing,
reach a bump with shallower slopes of $\approx -1.6\,(-1.8)$ at intermediate scales $k\approx 16\,(23)$ and get systematically steeper
in the dissipation range.\\
\item We interpret the bump in the slopes as numerical, non-physical bottleneck effect caused by the artificial numerical dissipation.
The bottleneck bump is more pronounced and appears on larger scales in the transverse spectra in comparison with the longitudinal spectra.\\
\item We find that the forcing method has a more dominant influence on the longitudinal spectra, such that the solenoidal forcing yields the same curved spectrum and
the compressive one yields a spectrum with a constant slope in the range $k \in (10:32)$ of $-1.94$ with the $2\sigma$-HDI $-1.95:-1.93$ and a standard deviation for
the time variations $\overline{\sigma_{\zeta}} = 0.04$.\\
\item We measured the variation of the slope $\zeta$ with time $\overline{\sigma_{\zeta}} \approx 0.1-0.2$ for a window size of $\Delta k = 6$,
which is independent of the forcing, $k$ scale, and the mode of the analysed spectra.
As observations measure only one time realisation of the power spectrum this uncertainty has to be taken into account.\\
\end{enumerate}

%%%%%%%%%%%%%%%%%%%%%%%
%%% APPENDIX          %
%%%%%%%%%%%%%%%%%%%%%%%
%\appendix
%\input{Appendix}

%%%%%%%%%%%%%%%%%%%%%%%
%%% ACKNOWLEDGMENTS   %
%%%%%%%%%%%%%%%%%%%%%%%
\section*{Acknowledgments}
 We thank Christoph Federrath, Brandon Kelly, Enrique Vazquez-Semadeni, and Stefanie Walch for stimulating discussions.
 L.K.~acknowledge financial support by the International Max
 Planck Research School for Astronomy and Cosmic Physics (IMPRS-A)
 and the Heidelberg Graduate School of Fundamental Physics
 (HGSFP). The HGSFP is funded by the Excellence Initiative of the
 German Research Foundation DFG GSC 129/1.
 L.K.,~R.S.K.~and R.S.~furthermore gives thanks for subsidies from
 the SFB 881 'The Milky Way System' (subproject B5, B2, \& B1) and via 
 the Deutsche Forschungsgemeinschaft (DFG) under grant KL 1358/11.
 R.S.K.~acknowledges support from the European Research Council under the European Community's Seventh Framework Programme (FP7/2007-2013) via the ERC Advanced Grant ``STARLIGHT:
 Formation of the First Stars'' (project number 339177).
 R.S.K.~thanks for the warm hospitality at the Kavli Institute for Particle Astrophysics and Cosmology at Stanford University and at the Department
 of Astronomy and Astrophysics at the University of California at Santa Cruz during a sabbatical stay in 2014 and 2015.
 P.G.~acknowledges the support by the Max-Planck Institut f\"{u}r Astrophysik in Garching,
 support from the DFG Priority Program 1573 \textit{Physics of the Interstellar Medium} and 
 acknowledges the support of a Marie Curie Research Training Network (MRTN-CT2006-035890).
 Supercomputing time at the Baden-W\"urttemberg cluster bwGRiD (http://www.bw-grid.de), member of the German D-Grid initiative,
 funded by the Ministry for Education and Research (Bundesministerium f\"ur Bildung und Forschung) 
 and the Ministry for Science, Research and Arts Baden W\"urttemberg (Ministerium f\"ur Wissenschaft, Forschung
 und Kunst Baden-W\"urttemberg) are gratefully acknowledged.
 The software used in this work was in part developed by the DOE-supported ASC / Alliance Center for Astrophysical
 Thermonuclear Flashes at the University of Chicago.\\

%%%%%%%%%%%%%%%%%%%%%%%
%%% BIBLIOGRAPHY      %
%%%%%%%%%%%%%%%%%%%%%%%
\bibliography{thesis}
%\bibliography{/home/lkonstandin/Dokumente/Publications/thesis}
\bibliographystyle{mn2e}

\label{lastpage}
\end{document}